\documentclass[a4paper,amsmath,amssymb,twocolumn,showpacs]{revtex4}
\usepackage{epsfig}
\usepackage{bm}

\newcommand{\vect}[1]{\mathbf{#1}}
\newcommand{\tdot}[1]{{\hskip2pt\ddot{\null}\hskip2.5pt
\dot{\null}\kern -5pt {#1}}}

\newcommand{\passo}{a}

\newcommand{\omegac}{\omega_c}
\newcommand{\omegap}{\omega_p}
\newcommand{\omegar}{\widetilde\omega}
\newcommand{\massae}{m_e}
\newcommand{\param}{\eta}
\newcommand{\paramc}{\eta_C}
\newcommand{\bparamc}{\eta^M}

\begin{document}

\title{A microscopic instability in neutral magnetized plasmas}


\author{M. Marino$^{1}$, M. Zuin$^{2}$, A. Carati$^{1}$, E. Martines$^{2}$ and L. Galgani$^{1}$}
\affiliation{$^{1}$Dipartimento di Matematica, Universit\`a degli
Studi di Milano, via Saldini 50, I-20133 Milano,
Italy\\$^{2}$Consorzio RFX, Associazione EURATOM-ENEA sulla
Fusione, Corso Stati Uniti 4, I-35127 Padova, Italy}

\date{June 23, 2010}
\begin{abstract}
We show that in a neutral magnetized plasma there exist
microscopic oscillatory modes, with wavelengths of the order of
magnitude of the mean interparticle distance, which become
unstable when the electron density exceeds a limit proportional to
the square of the magnetic field. The model we consider is just a
linearization of the classical one for neutral plasmas, namely a
system of electrons subjected to Coulomb interactions among
themselves and with a uniform positive neutralizing background.
This model is here dealt with as an actual many-body problem,
without introducing any averaging over the individual particles.
The expression of the density limit coincides, apart possibly from
a numerical factor of order one, with the well-known Brillouin
density limit for a nonneutral plasma, which has however a
macroscopic origin. The density limit here found has the same
order of magnitude as the operational density limit observed in
several conventional tokamak devices. We finally show that, when
the full electromagnetic interactions are taken into account,
dispersion relations are obtained which for short wavelengths
reduce to those obtained here in the purely Coulomb case, and for
long wavelengths reproduce the familiar ones of MHD.
\end{abstract}


\maketitle

\section{Introduction}
It is well known that the magnetic confinement of a pure electron
plasma is possible only for electron densities $n_e$ below the
so-called Brillouin density limit \cite{brill, davidson} given by
\begin{equation}\label{dl}
n_e^M= \bparamc\frac{\epsilon_0 B^2}{m_e}\,,
\end{equation}
where $\bparamc=1/2$, while $B$, $\epsilon_0$ and $m_e$ are the
magnetic field, the permittivity of free space, and the electron
mass. The condition $n_e<n_e^M$ can be equivalently expressed as
$\param<\bparamc$ in terms of the dimensionless parameter
\begin{equation}\label{parametro}
\param\equiv \frac {\massae
n_e}{\epsilon_0 B^2}=\frac{\omegap^2}{\omegac^2}\, ,
\end{equation}
where $\omegap= e \sqrt{n_e/ \epsilon_0\massae}$ and $\omegac=
Be/m_e$ are the electron plasma frequency and cyclotron frequency
respectively, $-e$ being the electron charge. The Brillouin
density limit for a nonneutral plasma is usually derived by
studying under a mean-field approximation the motion of a single
generic electron of the plasma: for densities beyond the limit,
the electrostatic repulsion due to the other charges cannot be
counterbalanced by the Lorentz confining force exerted by the
magnetic field.

In the present paper we show that a density limit exactly of the
form (\ref{dl}) --- possibly with a slightly different value of
the critical parameter $\bparamc$ --- exists for neutral plasmas
too, inasmuch as the plasma presents an internal dynamical
instability beyond that limit. Here however the instability can be
revealed only by dealing with a microscopic many-body model
involving the mutual Coulomb interaction of the individual
charges, because it mainly concerns modes of wavelength of the
order of the interparticle distance.

The model we consider is just the classical one of point electrons
obeying Newton's equations in an external magnetic field, with
Coulomb interactions both among themselves and with a smeared-out
positive neutralizing background. Such a model is considered for
example in the works of Bohm, Gross and Pines \cite{bg1, bp1,
pb2}, and in the previous ones of Langmuir and Tonks
\cite{langmuir, tonks}. In the impossibility of dealing with the
general analytical solution for such a many-body problem, those
authors introduced in the equations of motion some averaging with
respect to the individual particle positions, and as a consequence
were compelled to consider only plasma oscillations with
wavelengths much longer than the mean interparticle distance.
Since, on the contrary, we are interested in the study of modes
with short wavelengths, we have to adopt a different approach. We
thus choose to stick with the original many-body problem, and to
retain in the description of the system the microscopic
coordinates of all the individual electrons. In order to simplify
the mathematical equations, we then perform a linearization about
an equilibrium configuration of the system. As already pointed out
in a classical paper by Langmuir \cite{langmuir}, the equilibrium
condition requires the electrons to lie on the sites of some
regular lattice. It turns out that the normal mode solutions of
the linearized equations can be studied analytically, leading to
dispersion relations which depend parametrically on the plasma
density. If one considers in particular oscillations about a
simple cubic lattice it turns out that, for densities larger than
a limit of the form (\ref{dl}), there exists a relevant fraction
of modes for which the frequency becomes complex, and so the
system becomes unstable. This instability concerns modes of
wavelength of the order of the interparticle distance, and so
cannot be revealed by the methods of Bohm, Gross and Pines, or by
the equations of magnetohydrodynamics (MHD).

The essential microscopic nature of the phenomenon discussed here
also emerges from considerations of an energetic type. The origin
of the instability lies in the fact that the equilibrium
configuration here considered is not a minimum of the potential
energy of the system. Indeed, it will be shown that the potential
energy decreases for global displacements involving all the
electrons of the system, when such displacements are described by
plane waves with wavevectors having certain directions. In the
absence of a magnetic field, the normal modes associated with such
wavevectors are thus unstable. For a fixed density, an external
magnetic field of large enough magnitude can stabilize them, but
below a critical value the modes of shortest wavelengths along
those directions become unstable.

We will also briefly discuss the possible significance of this
instability in connection with the density limit empirically
encountered in the operation of tokamaks for fusion research on
magnetic confinement, pointing out that in several cases the
density limit found here is of the same order of magnitude as the
experimental one.

The equations of motion and their linearization are given in
section \ref{model}, together with the equation for the normal
modes and the corresponding expression of the energy. In section
\ref{sw} the form of the dispersion relations is studied in
dependence of the relevant parameter $\eta=\omegap^2/\omegac^2$,
and the existence of the instability is exhibited. In section
\ref{lw} a generalization of the model is considered, in which the
full electromagnetic interactions are introduced, including
retardation and radiative terms. We prove that the purely Coulomb
model considered in the previous sections is recovered for short
wavelengths, while in the long wavelength limit the dispersion
relations exactly coincide with those provided by the macroscopic
equations of MHD for a low temperature plasma. Finally, the
possible physical relevance of the instability discussed here is
briefly addressed in the Conclusion.

Three appendices are devoted to the technical details of some
calculations. They concern respectively the effective field acting
on an electron inside the plasma, the electrostatic energy in the
equilibrium configuration of the plasma, and the dependence of the
instability threshold on the orientation of the magnetic field.

\section{The model}\label{model}

\subsection{The equations of motion and their linearization}

Denoting by $\vect z_i$ the position vector of the $i$-th
electron, its equation of motion is
\begin{equation}\label{moto}
m_e \ddot {\vect z}_i= -e {\vect e}_i({\vect z}_i, t) -e \dot
{\vect z}_i \times {\vect B} \,,
\end{equation}
where ${\vect e}_i$ is the Coulomb field generated by all the
other electrons and by the positive background, and ${\vect B}$ is
an external magnetic field, which is supposed to be constant.

In order to obtain a linearized system of equations of motion, we
look for an equilibrium configuration for the electrons. If the
plasma is assumed to be infinite (i.e., if all edge effects are
neglected), it is easy to see that such an equilibrium
configuration is given by the points of any arbitrary
simple Bravais lattice. 
Naming ${\bf a}_{1}$, ${\bf a}_{2}$, ${\bf a}_{3}$ the primitive
translation vectors of the lattice, we have $n_e=1/V$, where
$V=|{\bf a}_{1}\cdot {\bf a}_{2}\times {\bf a}_{3}|$ is the volume
of the primitive cell, so that $\passo=V^{1/3}$ represents the
lattice parameter. We denote by ${\bf r}_{{\bf n}}= n_{1}{\bf
a}_{1}+ n_{2}{\bf a}_{2} +n_{3}{\bf a}_{3}$ the position vector of
an arbitrary point of the lattice, labelled by the triple of
relative integers $\mathbf{n}=(n_1, n_2, n_3) \in \mathbf{Z}^3$.
We shall also label with ${\vect n}$ the electron associated with
this lattice site, and so we denote by ${\vect z}_{\vect n}$ its
position vector. Finally, we introduce the corresponding
displacement ${\vect x}_{\vect n}$ by
\[
{\vect z}_{\vect n}(t)={\vect r}_{\vect n}+ {\vect x}_{\vect
n}(t)\,.
\]

The linearized equations of motion about the chosen equilibrium
configuration of the system can be shown, for all $\vect n\in
\vect Z^3$, to be
\begin{equation}\label{motolin}
m_e \ddot {\vect x}_\mathbf{n}= - \frac{n_e e^2}{3 \epsilon_0}
{\vect x}_\mathbf{n} - e\dot {\vect x}_\mathbf{n} \times {\vect B}
+\frac{e^2}{4\pi \epsilon_0} \sum_{{\vect m}\neq {\vect n}}
\widehat{\vect D}_{\vect n -\vect m}\cdot \vect x_{\vect m}\,,
\end{equation}
where $\widehat{\vect D}_{\vect m}$, for $\vect m \in \vect Z^3$,
$\vect m \neq \vect 0$, is a symmetric matrix whose elements
depend on the lattice geometry, namely
\begin{equation}\label{d}
(\widehat{\vect D}_{\vect m})_{ij} = \bigg[3\frac{({\vect
r}_{\mathbf{m}})_i ({\vect r}_{\mathbf{m}})_j}{|{\vect
r}_{\mathbf{m}}|^5} -\frac{\delta_{ij}}{|{\vect
r}_{\mathbf{m}}|^3}\bigg]\,.
\end{equation}

In order to prove (\ref{motolin}), let us start from the general
equation (\ref{moto}), and observe that up to first order in the
displacements ${\vect x}_\mathbf{m}$ of the electrons one can
write
\[
{\vect e}_{\vect n}({\vect z}_{\vect n}, t)= {\vect e}_{\vect
n}^{(0)}({\vect z}_{\vect n})+ {\vect e}_{\vect n}^{(1)}({\vect
r}_{\vect n}, t)\,,
\]
where ${\vect e}_{\vect n}^{(0)}$ and ${\vect e}_{\vect n}^{(1)}$
are respectively the contributions to the field ${\vect e}_{\vect
n}$ of order zero and one in the ${\vect x}_\mathbf{m}$.

It is clear that ${\vect e}^{(0)}_{\vect n}$ is given by the
constant electrostatic field generated by the background and by
the electrons ${\vect m}\neq {\vect n}$, when all these electrons
are kept fixed at their equilibrium positions. Since the Bravais
lattice is invariant under spatial reflections, we have ${\vect
e}^{(0)}_{\vect n}({\vect r}_{\vect n})=0$. This is actually the
reason why, as we said before, the points of the lattice are
equilibrium positions for the electrons. Moreover, since the
plasma is assumed to be globally neutral, the charge density of
the background must be $\rho_{\rm bg} =n_e e$. From $\textrm
{div}\, {\vect e}^{(0)}_{\vect n}({\vect r}_{\vect n}) = \rho_{\rm
bg}/ \epsilon_0 = n_e e/\epsilon_0$, assuming that the lattice is
isotropic it then follows that $\partial ({\vect e}^{(0)}_{\vect
n})_i /\partial x_j ({\vect r}_{\vect n})= \delta_{ij} n_e
e/3\epsilon_0$. Hence, to first order in ${\vect x}_{\vect n}$ we
can write
\begin{equation}\label{e0}
{\vect e}^{(0)}_{\vect n}({\vect z}_{\vect n})= \frac{n_e e}{3
\epsilon_0} {\vect x}_\mathbf{n}\,.
\end{equation}

Concerning ${\vect e}_{\vect n}^{(1)}$, it is given by the sum of
the Coulomb fields generated by all the electrons ${\vect m}\neq
{\vect n}$, computed at order one in their displacements ${\vect
x}_{\vect m}$. Such contributions are given by the well-known
expression (see for instance chapter 4 of ref.\ \cite{Jackson}) of
the field $\vect E_{\vect m}$ generated by an electric dipole
$-e{\vect x}_{\vect m}$ located at $\vect r_{\vect m}$, that is
\begin{equation}\label{dipc}
\vect E_{\vect m}(\vect x) = -\frac{e}{4\pi \epsilon_0}
\left(3\frac {\vect x_{\vect m}\cdot \vect y} {y^5}\vect y -
\frac{\vect x_{\vect m}}{y^3}\right)\,,
\end{equation}
with $\vect y=\vect x- \vect r_{\vect m}$. It follows that
\begin{equation}\label{dip}
{\vect e}_{\vect n}^{(1)} ({\vect r}_{\vect n},t)= \sum_{{\vect
m}\neq {\vect n}} \vect E_{\vect m}(\vect r_{\vect n})=
-\frac{e}{4\pi \epsilon_0} \sum_{{\vect m}\neq {\vect n}}
\widehat{\vect D}_{\vect n -\vect m}\cdot \vect x_{\vect m}\,,
\end{equation}
with $\widehat{\vect D}$ given by (\ref{d}). From (\ref{e0}) and
(\ref{dip}), equation (\ref{motolin}) is then readily obtained.

\subsection{The equation for the normal modes}\label{nm}

In order to deal with the infinite system of linear differential
equations (\ref{motolin}), we shall look as usual for normal mode
solutions of the form
\begin{equation}\label{normmode}
{\vect x}_{{\vect n}}={\vect C}\exp \left[ i({\vect k} \cdot {\bf
r}_{{\bf n}}-\omega t)\right]\,,
\end{equation}
where the wavevector ${\vect k}$, the frequency $\omega$ and the
polarization vector ${\vect C}$ are constants. For such an ansatz,
the field ${\vect e}_{\vect n}^{(1)} ({\vect r}_{\vect n}, t)$
becomes
\begin{equation}\label{e1}
{\vect e}_{\vect n}^{(1)} ({\vect r}_{\vect n}, t) =- \frac {e
n_e}{\epsilon_0} \widehat {\vect M}({\vect q}) \cdot \vect C
\exp[i(\vect k \cdot \vect r_{\vect n}-\omega t)]\,,
\end{equation}
where we have introduced the real dimensionless symmetric matrix
\begin{equation}\label{mij0}
\widehat {\vect M}({\vect q})= \frac V {4\pi}\sum_{{\vect m}\neq
{\vect 0}} \widehat{\vect D}_{\vect m}\exp(2\pi i\vect q \cdot
\vect r_{\vect m}/a)\,,
\end{equation}
which depends on the wavevector $\vect k$ and on the lattice
parameter $a$ only through their product
$$
\vect q \equiv \vect k a/2\pi\,.
$$
Note that the series in
(\ref{mij0}) converges only in an improper sense. In appendix
\ref{appa}, using techniques analogous to those already developed
in \cite{mcg}, we obtain an expression for $\widehat {\vect
M}({\vect q})$ given by the sum of an absolutely convergent
series. For an isotropic lattice, this expression can be written
as
\begin{equation}\label{mij}
M_{ij}({\vect q})=\frac {\delta_{ij}}3  -\frac{q_i q_j} {{\vect
q}^2}+ N_{ij}({\vect q})\,,
\end{equation}
with
\begin{align}\label{nij}
N_{ij}({\vect q})=\ &(\alpha/3-2\beta){\bf q}^{2}\delta_{ij} +
(\alpha- 4\beta)q_{i}q_{j} \nonumber\\
&+ 2(5\beta- \alpha) q_{i}^{2} \delta_{ij} - \sum_{\vect m \neq 0}
\bar c_{ij}(\vect H_{\vect m}, \vect q) \,.
\end{align}
Here the sum runs over all the points of the dimensionless
reciprocal lattice with $\vect m \neq 0$, namely
\begin{equation}\label{hm}
\vect H_{\vect m}\equiv \frac 1{a^2} (m_1 \vect a_2 \times \vect
a_3+ m_2 \vect a_3 \times \vect a_1+ m_3 \vect a_1 \times \vect
a_2)\,,
\end{equation}
and the function $\bar c_{ij}$ is obtained by subtracting to the
function
\begin{equation}\label{cij}
c_{ij}(\vect H, \vect q)=\frac{(H_{i} +q_{i}) (H_{j} +q_{j})}
{({\bf H}+ {\bf q})^{2}}
\end{equation}
the terms of order $k\leq 3$ of its Taylor expansion in the
variable $\vect q/H$ about the origin. Finally, the constants
$\alpha$ and $\beta$ can be numerically computed for any given
geometry of the lattice using formulas
(\ref{alpha})--(\ref{beta}). Note that ${\rm Tr}\, \widehat {\vect
N} (\vect q)=0$ for all $\vect q$.

The term $\delta_{ij}/3$ on the right-hand side of (\ref{mij}) is
the equivalent of the so-called ``Lorentz term'' in the expression
of the local field inside isotropic dielectrics. Its contribution
to the equation of motion exactly cancels the first term on the
right-hand side of (\ref{motolin}).

In conclusion, the normal mode ansatz (\ref{normmode}) for the
linearized equation of motion (\ref{motolin}) leads to a linear
equation for the polarization vector $\vect C$, namely
\begin{equation}\label{clin}
i \frac {\omega e}{m_e}\vect B \times \vect C -\omega^2 \vect C
=\omegap^2 \left[-\frac{\vect q }{\vect q^2} \vect q \cdot \vect
C+\widehat {\vect N}({\vect q}) \cdot \vect C \right]\,,
\end{equation}
where $\omegap= e \sqrt{n_e/ \epsilon_0\massae}$ is the electron plasma
frequency. This is an  equation of the form $\widehat {\vect A}
\cdot\vect C=0$, with the matrix $\widehat {\vect A}$ given by
\begin{equation}\label{amat}
A_{ij}(\vect q, \omega)=\ i  \frac {\omega e}{m_e}
\epsilon_{ijk}B_k+ \omega^2 \delta_{ij} -\omegap^2 \left[\frac{q_i
q_j}{q^2} -N_{ij} ({\vect q})\right]\,,
\end{equation}
where $\epsilon_{ijk}$ denotes the completely antisymmetric tensor
such that $\epsilon_{123}=1$. The condition for the existence of
nontrivial solutions is ${\rm det}\, \widehat {\vect A} (\vect q,
\omega) =0$. By solving this last equation with respect to
$\omega$ for a given $\vect q$, one obtains the dispersion
relation for the oscillations in our model of magnetized plasma.

\subsection{The energy}\label{energy}
As the normal modes discussed here have a purely electrostatic
nature, it is of interest to have available an analytical
expression for the electrostatic energy $U$ of the system. It is
easy to see that
\begin{equation}\label{u1}
U=-\frac e2 \sum_{\vect n} \phi_{\vect n}(\vect z_{\vect n})\,,
\end{equation}
where $\phi_{\vect n}$ is the electrostatic potential generated by
all the charges of the plasma (electrons and background) except
the electron $\vect n$. The energy $U_0$, corresponding to the
configuration in which all the electrons are at their equilibrium
positions, i.e.\ $\vect z_{\vect n}= \vect r_{\vect n}$ for all
$\vect n \in \vect Z^3$, can be evaluated by means of a suitable
modification of the Ewald method for the calculation of the
electrostatic energy of a ionic lattice (see for instance appendix
B of \cite{kittel}). If $N$ is the total number of electrons in
the plasma, which is assumed to be so big that the surface effects
can be neglected, we have
\begin{equation}\label{u0}
U_0= -N\frac{b e^2}{8\pi \epsilon_0 a}\,,
\end{equation}
where the dimensionless constant $b$ can be calculated using the
formula
\begin{align}\label{mu}
b= &-\sum_{\vect m\neq \vect 0} \frac{\exp(-\pi \vect H_{\vect
m}^2/\xi)}{\pi\vect H_{\vect m}^2} -\sum_{\vect n \neq \vect
0}\frac{F(\sqrt{\pi\xi} |\vect r_{\vect n}|/a)}{|\vect r_{\vect
n}|/a}\nonumber \\
&+\frac 1\xi +2\sqrt{\xi}\,,
\end{align}
involving a positive parameter $\xi$. In this formula the function
$F$ is defined as
\begin{equation}\label{funf}
F(x)= \frac 2\pi \int_{x}^{+\infty}e^{-s^2}ds \,.
\end{equation}
It can be proved that the right-hand side of (\ref{mu}) is
independent of the value of $\xi$, and that both sums converge
very quickly for $\xi$ of order unity. The calculations leading to
(\ref{mu}) are carried out in appendix \ref{appb}.

Since ${\vect e}_{\vect n}({\vect z}_{\vect n}, t)=-{\mathbf
\nabla}\phi_{\vect n}(\vect z_{\vect n})$, it follows from
(\ref{e0}) and (\ref{dip}) that up to second order in the
displacements $\vect x_{\vect n}$ we have
\begin{equation}\label{u}
U= U_0+ \frac{n_e e^2}{6 \epsilon_0} \sum_{\vect n}{\vect
x}_\mathbf{n}^2- \frac{e^2}{8\pi \epsilon_0} \sum_{{\vect m}\neq
{\vect n}} \vect x_{\vect n} \cdot \widehat{\vect D}_{\vect n
-\vect m}\cdot \vect x_{\vect m}\,.
\end{equation}
The total energy, which is conserved on the solutions of the
equations of the motion, is then $E=T+U$, where
$T=(m_e/2)\sum_{\vect n} \dot{\vect x}_{\vect n}^2$ is the kinetic
energy.

For a normal mode of the form (\ref{normmode}), it follows from
(\ref{u}) that the electrostatic energy per electron $U/N$ is
given by
\begin{align}\label{uq}
\frac U N= &-\frac{b e^2}{8\pi \epsilon_0 a}+ \frac{m_e
\omegap^2}{4} \vect C^* \cdot \left[\hat{\vect 1}/3- \hat{\vect
M}(\vect q)\right] \cdot \vect C \\
= &-\frac{b e^2}{8\pi \epsilon_0 a}+ \frac{m_e \omegap^2}{4}
\left[\frac{|\vect q \cdot \vect C|^2}{\vect q^2}- \vect C^*
\cdot\hat{\vect N}(\vect q)\cdot \vect C\right] \,. \nonumber
\end{align}
Then, for the total energy per electron $E/N$ of the normal mode
we obtain
\begin{align*}
\frac E N= &-\frac{b e^2}{8\pi \epsilon_0 a}+ \frac{m_e
\omegap^2}{4} \vect C^* \cdot \left[\hat{\vect 1}/3- \hat{\vect
M}(\vect q)\right] \cdot \vect C \\
&+ \frac{m_e \omega^2}{4} |\vect C|^2\\
=&-\frac{b e^2}{8\pi \epsilon_0 a}+ \frac{m_e \omega^2}{2} |\vect
C|^2 -i\frac{\omega e}4\vect B\cdot \vect C \times \vect C^*  \,,
\end{align*}
where the last equality follows from the equation of motion
(\ref{clin}).

Equation (\ref{u}) shows that the potential energy always
increases when a single electron is displaced from its equilibrium
position, all the others being kept fixed. We see however from
(\ref{uq}) that, if for some $\vect q$ the matrix $\omegap^2
(\hat{\vect 1}/3- \hat{\vect M}(\vect q))$ has a negative
eigenvalue corresponding to some eigenvector $\vect C$, then the
potential energy is decreased by a simultaneous displacement of
all the electrons according to the pattern described by the
polarization vector $\vect C$ and the wavevector $\vect q$. This
means that, in such a case, the potential energy at the
equilibrium configuration does not present a minimum. This is
directly connected with the existence of unstable modes since,
according to the dynamical equation (\ref{clin}), for $\vect B=0$
the squared frequency $\omega^2$ of the normal modes is just given
by an eigenvalue of the matrix $\omegap^2(\hat{\vect 1}/3-
\hat{\vect M}(\vect q))$. Hence negative eigenvalues give rise to
imaginary values of the frequency. We will see in the next section
that negative eigenvalues actually exist in the case of the simple
cubic lattice, when $\vect q$ is parallel to one of the three
principal axes.

\section{Unstable modes of oscillation}\label{sw}

\subsection{Dispersion relations and existence of unstable
modes}\label{sw1}

To find the dispersion relations we have to solve the linear
equation (\ref{clin}). This contains the function ${\vect
N}({\vect q})$ given by (\ref{nij}), which depends on the
particular geometry of the Bravais lattice. In the present paper
we limit ourselves to considering the easiest possible case,
namely that of a simple cubic lattice, for which the primitive
translation vectors are parallel to the three basic unit vectors
of a cartesian frame: $\vect a_i= a\vect u_i$, $i=1,2,3$. In this
case the reciprocal lattice is also simple cubic, and we have from
(\ref{hm}) $\vect H_{\vect m}= \vect m$, with ${\vect m}\in \vect
Z^3$. For this lattice the geometrical constants appearing in
(\ref{nij}) are $\alpha\cong 8.9136$ and $\beta \cong 1.2267$.
Moreover, the constant $b$ appearing in the expression (\ref{u0})
of the electrostatic energy at equilibrium is $b\cong 2.8373$.

\begin{figure}
\includegraphics[width=9cm]{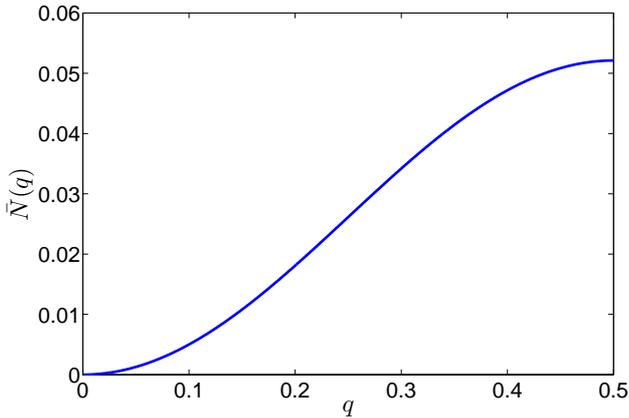}
\caption{Graph of the function $\bar N(q)$.} \label{fig2}
\end{figure}

Let us consider the particular case in which the wavevector $\vect
k$ is parallel to a principal lattice axis, say $\vect u_3$. In
such a case the matrix ${\vect N}({\vect q})$ becomes diagonal. We
shall denote $N_{11}(q{\vect u}_3)= N_{22}(q{\vect u}_3)\equiv
\bar N(q)$, so that $N_{33}(q{\vect u}_3)= -2\bar N(q)$. The graph
of the function $\bar N(q)$ is displayed in Fig.\ \ref{fig2}. It
is easily seen from (\ref{clin}) that the dispersion relations,
when expressed in terms of the quantities $q$ and
\[
\omegar\equiv \omega/\omegac\,,
\]
contain the single positive parameter $\param=
\omegap^2/\omegac^2$. For instance, if $\vect B$ is also parallel
to $\vect u_3$, then the dispersion relation for transversal
normal modes is implicitly expressed by the equation
\begin{equation}\label{bp}
\omegar^2- \omegar +\param \bar N(q) = 0 \,.
\end{equation}
For these modes, the electrons move along circular orbits in
planes orthogonal to $\vect B$.

\begin{figure}
\includegraphics[width=9cm]{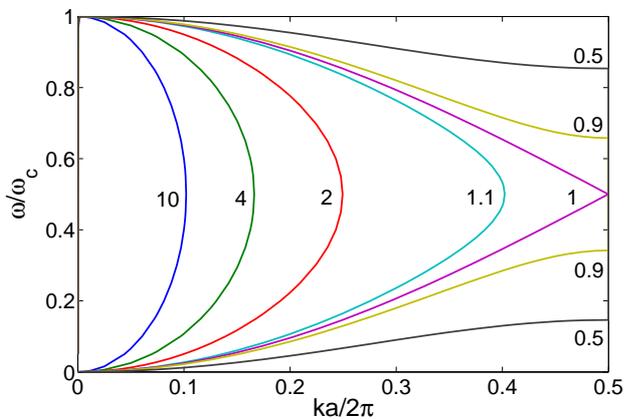}
\caption{Dispersion curves for $\vect k$ and $\vect B$ both
parallel to an axis of the simple cubic lattice, and circular
transversal polarization. The number next to each curve gives the
corresponding value of the ratio $\param/\paramc$, with
$\param=\omegap^2/\omegac^2$ and $\paramc\cong 4.80$.}
\label{fig1}
\end{figure}

In Fig.\ \ref{fig1} we plot the dispersion curves obtained from a
numerical solution of (\ref{bp}) for various values of $\param$.
This is the figure in which the phenomenon of microscopic plasma
instabilities manifests itself. We see in fact that the curves are
defined in the whole Brillouin zone $0\leq q \leq 1/2$ only for
$\param$ below a certain critical value $\paramc\cong 4.80$. This
obviously follows from the fact that (\ref{bp}) is a second degree
equation in $\omegar$, which admits two real solutions only for
$\param\leq 1/(4 \bar N(q))$. For $\param=\paramc$ there exists a
double real solution for $q=1/2$, hence
\[
\paramc= \frac 1 {4 \bar N(1/2)}\cong 4.80\,.
\]
If $\param>\paramc$, for $q$ sufficiently near to 1/2 (i.e.\ for
sufficiently short wavelengths) the two solutions of (\ref{bp})
are complex conjugate. For one of these solutions the electrons
simultaneously spiral away from their equilibrium positions, so
that the normal mode becomes unstable.

Considering again the case in which both $\vect k$ and $\vect B$
are parallel to $\vect u_3$, from (\ref{clin}) we have for the
longitudinal modes
\begin{equation}\label{long}
\omegar^2= \param\left(1+2 \bar N(q)\right)\,.
\end{equation}
Since the right-hand side of this equality is always positive, we
see that for these modes $\omega$ is real for all $q$ and all
$\param$.

Let us now consider the case in which $\vect k$ is still parallel
to $\vect u_3$, but $\vect B$ is parallel to $\vect u_2$. For the
transversal mode with $\vect C$ parallel to $\vect B$ we find the
relation $\omega^2= -\omegap^2\bar N(q)$, or
\begin{equation}
\omegar^2= -\param\bar N(q) \,.
\end{equation}
The right-hand side is in this case always negative, hence this
equation provides two opposite imaginary values for $\omega$. This
means that, for all $\param
>0$, there exists an unstable mode for which the electrons move
exponentially away from the equilibrium positions along the
direction of $\vect B$. Furthermore, for elliptic orbits
orthogonal to $\vect B$ we have
\begin{equation}\label{ellipt}
\omegar^4- \left[1+ (1+\bar N(q)) \param\right]\omegar^2 -\bar
N(q)(1+2\bar N(q))\param^2 = 0 \,.
\end{equation}
This second degree equation in $\omegar^2$ always admits two real
solutions of opposite signs, hence there exist a stable and an
unstable mode of this type for all $q$ and all $\param>0$.

\begin{figure*}
\includegraphics[width=19cm]{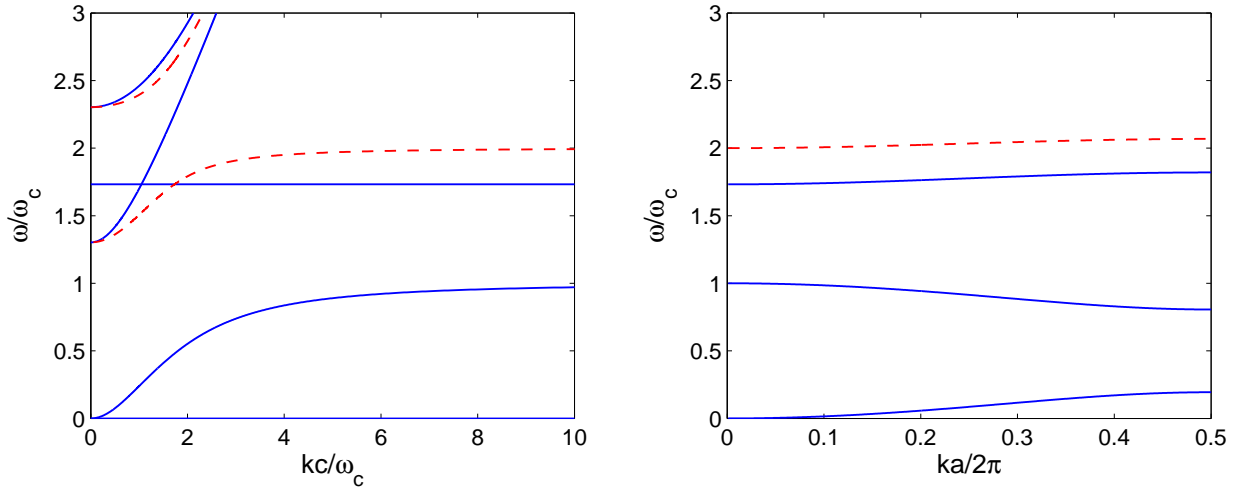}
\caption{Dispersion curves obtained from our model for
$\omegap^2/\omegac^2=3$, at two different scales of the $k$ axis.
It is assumed that $\omegac a/2\pi c \approx 10^{-5}$, so the left
graph shows the behavior for $ka/2\pi\lesssim 10^{-4}$ of the
curves of the right graph. The left graph (see section \ref{lw})
is independent of the lattice structure, whereas the right graph
is specific of the modes propagating along an axis of a simple
cubic lattice. Solid lines describe modes with $\vect B$ parallel
to $\vect k$, dashed lines modes with $\vect B$ perpendicular to
$\vect k$.} \label{fig0}
\end{figure*}

The dispersion curves for all the stable modes considered above,
with $\vect q$ parallel to a lattice axis, are reported in the
right part of Fig.\ \ref{fig0} for the particular value
$\param=3$. The solid lines correspond to the cases in which
$\vect B$ is parallel to $\vect q$. In particular, the two lower
ones refer to transversal modes and are derived from (\ref{bp}).
They can thus be compared with the curves of Fig.\ \ref{fig1},
noticing that $\param=3$ corresponds to $\param/\paramc=0.625$.
The top solid line refers instead to the longitudinal modes
described by (\ref{long}). Finally, the dashed line corresponds to
the case in which $\vect B$ is orthogonal to $\vect q$, and is
derived from (\ref{ellipt}).

The left part of Fig.\ \ref{fig0} shows the behavior of the same
curves in the long-wavelength limit, i.e.\ for $q\ll 1$, as
resulting from the full electromagnetic treatment given in section
\ref{lw}. We will show that in this limit the dispersion relations
derived from our model exactly coincide with those provided by the
equations of MHD for a zero temperature plasma.

\subsection{Estimate of the critical parameter
$\bparamc$}\label{est}

\begin{figure}
\includegraphics[width=9cm]{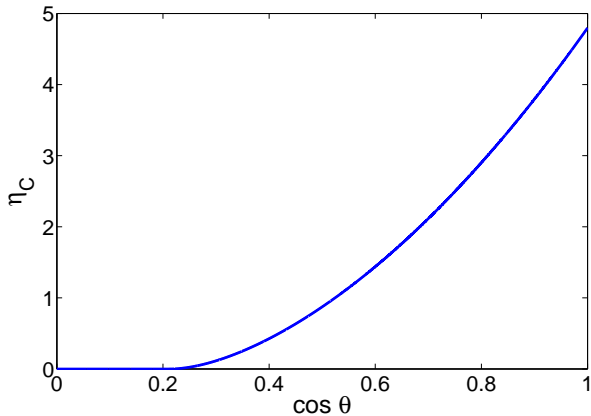}
\caption{Graph of the function $\paramc$ versus $\cos\theta$.}
\label{fig3}
\end{figure}

From the cases just considered it appears that the threshold
$\paramc$ for the onset of plasma instabilities depends on the
angle $\theta$ between $\vect B$ and $\vect q= \vect k a/2\pi$. We
can thus write $\paramc=\paramc (\cos\theta)$, with $\paramc(0)=
0$, $\paramc(1)\cong 4.80$. For a generic $\theta$, since $\bar N$
is an increasing function of $q$, the instability will first
manifest itself at the edge $q=1/2$ of the Brillouin zone. Hence,
to determine $\paramc(\cos\theta)$ we look for the solutions
$\omega$ of the equation ${\rm det}\, \widehat {\vect A} (\vect k,
\omega) =0$, where $\widehat {\vect A} (\vect k, \omega)$ is given
by formula (\ref{amat}) for $\vect q= (1/2) \vect u_3$. The
critical value $\paramc(\cos\theta)$ is determined as the largest
$\param$ for which all these solutions are real (see appendix
\ref{theta} for the details of the calculation). The graph of the
function $\paramc (\cos\theta)$ obtained in this way is shown in
Fig.\ \ref{fig3}.

Since there is no a priori correlation between the direction of
$\vect B$ and the orientation of the cubic lattice, it seems
reasonable to associate to our model of plasma the critical
parameter $\bparamc$ which is obtained by averaging the function
$\paramc (\cos\theta)$ over the full solid angle. One thus finds
with a numerical integration
\[
\bparamc = \int_0^1 \paramc (\cos\theta) d(\cos \theta) \cong
1.40\,.
\]
Recalling the definition (\ref{parametro}) of the parameter
$\param$, we conclude that to any given value of $B$ one can
associate a critical value $n_e^M$ of the electronic density given
by (\ref{dl}), with $\bparamc\cong 1.40$. For densities above this
threshold, unstable normal modes are expected to arise within the
plasma in a significant way.

By means of a more extensive study of the behavior of the matrix
$\widehat {\vect M}(\vect q)$ as a function of $\vect q$, it is
possible to show that there exist also unstable modes for which
the wave propagates along other lattice directions. However, the
case considered in this section, for which $\vect q$ is parallel
to a lattice axis, is the most significant one, since for any such
$\vect q$ there are two independent transversally polarized
unstable modes. Moreover, the instability of these modes can be
removed by the presence of a suitable external magnetic field.
This is precisely the mechanism which is responsible for the
prediction of a density limit proportional to the squared magnetic
field.

Of course, we could have taken a different Bravais lattice,
instead of a simple cubic one, as the equilibrium configuration of
the electrons. In such a case, the number and the properties of
the unstable modes would in general have been different, since the
matrix $\widehat {\vect M}(\vect q)$ depends on the specific
geometry of the lattice. As a consequence, also the value of the
critical parameter $\bparamc$ is expected to be dependent on the
choice of the equilibrium lattice. However, a systematic
investigation of this dependence falls outside the scope of the
present work.

\section{The electrodynamical extension of the model}\label{lw}

\subsection{The equation for the normal modes}
The characteristic feature of the present approach consists in
linearizing the equations of motion of the classical plasma model
about an equilibrium configuration. By considering purely
coulombian interactions, we have exhibited the existence of
instabilities at short wavelengths which were not revealed by
other treatments of the model. In the present section we want to
show that, on the other hand, the present approach exactly
reproduces for long wavelengths the dispersion relations which are
usually obtained for cold plasmas by applying the continuum
equations of MHD or the methods of references \cite{bg1,bp1,pb2}.

To this end, we note that the analytical treatment of plasma
oscillations in the dipole approximation, which has been given in
section \ref{model} for purely coulombian interactions, can be
generalized in a straightforward way so as to make use of the full
electrodynamical expression of the field, thus taking into account
also radiative terms and retardation. This amounts to replacing
the electrostatical expression (\ref{dipc}) of the dipole field
with the complete one (see for instance chapter 9 of ref.\
\cite{Jackson})
\begin{align}\label{dipf}
\vect E_{\vect m}(\vect x, t)= &-\frac{e}{4\pi \epsilon_0} \bigg[
3\frac{({\vect x}_{\vect m}\cdot {\vect y}) {\vect y}}{y^5}-
\frac{{\vect x}_{\vect m}}{y^3} +3\frac{(\dot{\vect
x}_{\vect m}\cdot {\vect y}) {\vect y}}{cy^4}  \nonumber\\
&- \frac{\dot{\vect x}_{\vect m}}{cy^2}  + \frac{(\ddot{\vect
x}_{\vect m}\cdot {\vect y}) {\vect y}}{c^2 y^3} -
\frac{\ddot{\vect x}_{\vect m}} {c^2 y}\bigg]\,,
\end{align}
where $\vect y= \vect x -{\vect r}_{\mathbf{m}}$. In this formula
the vector ${\vect x}_{\vect m}$ and all its time derivatives are
evaluated at the retarded time $t_{\rm ret}=t-y/c$. The resulting
equations for a normal mode essentially coincide with those
obtained in reference \cite{mcg}, so we will here limit ourselves
to briefly recalling the results. By summing the retarded fields
$\vect E_{\vect m}$ generated by all the individual electrons
$\vect m \neq \vect n$, we obtain in place of (\ref{e1})
\begin{align}\label{e1rad}
{\vect e}_{\vect n}^{(1)} ({\vect r}_{\vect n}, t) =&\left[- \frac
{e n_e}{\epsilon_0} \widehat {\vect M}({\vect q}, f) \cdot {\vect
C}+\frac{ie\omega^3}{6\pi \epsilon_0 c^3}
{\vect C} \right]\nonumber \\
&\times \exp \left[ i({\vect k} \cdot {\bf r}_{{\bf n}}-\omega
t)\right]\,,
\end{align}
where $\widehat {\vect M}({\vect q}, f)$ is a dimensionless
symmetric matrix which depends on the rescaled wavevector $\vect q
\equiv \vect k a/2\pi$ and the rescaled frequency $f\equiv \omega
a/2\pi c$. A remarkable fact is that $\widehat {\vect M}({\vect
q}, f)$ is real when its arguments $\vect q$  and $f$ are real.
Moreover, the term $(ie\omega^3/6\pi \epsilon_0 c^3) {\vect C}$
inside the square brackets of (\ref{e1rad}) is exactly cancelled
by the term $(e^2/6\pi \epsilon_0 c^3)\, \dddot {\vect
z}_\mathbf{n}$ which, according to the Lorentz--Dirac equation
(see \cite{dirac, marino} and chapter 17 of \cite{Jackson}), has
to be added to the right-hand side of the equation of motion
(\ref{moto}) in order to take into account radiation reaction.
Comments on the profound mathematical and physical meaning of this
cancellation are given in references \cite{cg} and \cite{mcg}. It
has however to be noted that, in any case, the strength of the
radiation reaction is completely negligible as far as the study of
the dispersion relations in a plasma is concerned.

It is useful to decompose the matrix $\widehat {\vect M}$ as
\[
\widehat {\vect M}({\vect q}, f)= \widehat {\vect M}^{\rm
lw}({\vect q}, f)+ \widehat {\vect N}({\vect q}, f)\,,
\]
where $\widehat {\vect M}^{\rm lw}({\vect q}, f)$ is the dominant
term for long-wavelengths, i.e.\ for $q\ll 1$, while $\widehat
{\vect N}({\vect q}, f)$ is a short-wavelength term, which becomes
relevant only when $a$ is not negligible with respect to the
wavelength $\lambda=2\pi/k$. For an isotropic lattice one obtains
\begin{equation}\label{matm}
\widehat {\vect M}^{\rm lw}({\vect q}, f)= \widehat {\vect M}^{\rm
mac}({\vect k}, \omega)+ \frac {\widehat{\vect 1}}3 \,,
\end{equation}
where $\widehat {\vect 1}$ is the $3\times 3$ identity matrix, and
\[
M_{ij}^{\rm mac}({\vect k}, \omega)= \frac{\omega^2
 \delta_{ij} - c^2 k_i k_j} {c^2{\vect k}^2- \omega^2}
= \frac{f^2 \delta_{ij}- q_i q_j} {{\vect q}^2- f^2}
\]
is the term associated with the macroscopic field ${\vect E}^{\rm
mac}$ inside the plasma. In the long-wavelength limit we have in
fact
\[
{\vect E}^{\rm mac}(\vect x, t)= - \frac{e n_e} {\epsilon_0}
\widehat {\vect M}^{\rm mac} ({\vect k}, \omega) \cdot \vect C
\exp \left[ i({\vect k} \cdot \vect x-\omega t)\right]\,.
\]
We have finally for the short-wavelength part:
\begin{align}\label{msw}
N_{ij}({\vect q}, f)=\ &[(\alpha/3-2\beta){\bf
q}^{2}-(2\alpha/3) f^2] \delta_{ij} \nonumber\\
&+ (\alpha- 4\beta)q_{i}q_{j} + 2(5\beta- \alpha) q_{i}^{2}
\delta_{ij} \nonumber\\
&- \sum_{\vect m \neq 0} \bar c_{ij}(\vect H_{\vect m}, \vect q,
f) \,,
\end{align}
where the function $\bar c_{ij}$ is obtained by subtracting to the
function
\begin{equation*}
c_{ij}(\vect H, \vect q, f)=\frac{(H_{i} +q_{i}) (H_{j} +q_{j})
-f^{2}\delta_{ij}} {({\bf H}+ {\bf q})^{2}- f^{2}}
\end{equation*}
the terms of order $k\leq 3$ of its Taylor expansion in the
variables $\vect q/H$ and $f/H$ about the origin. This implies
that in the long wavelength limit (i.e.\ $q \ll 1$, $f \ll 1$)
$\bar c_{ij}$ is of order four in the variables $q$ and $f$. It
follows that in this limit the leading term of $\widehat{\vect
N}({\vect q}, f)$ is represented by the second-degree homogeneous
polynomial appearing on the right-hand side of (\ref{msw}).

From these results, one deduces that the linearized dynamical
equation for a normal mode is
\begin{align}\label{clinrad}
&i \frac {\omega e}{m_e}\vect B \times \vect C -\omega^2 \vect C
\nonumber \\
=\ &\omegap^2 [\widehat {\vect M}^{\rm mac}({\vect k}, \omega)+
\widehat {\vect N}({\vect q}, f)] \cdot \vect C \,,
\end{align}
which represents the generalization of (\ref{clin}) when the full
electrodynamic interaction is taken into account.

\subsection{The Coulomb limit}
It is now interesting to establish under which conditions the
purely coulombian equation (\ref{clin}) represents a good
approximation of (\ref{clinrad}). We first note in this respect
that the proportionality factor $\beta$ between $f$ and
$\tilde\omega = \omega/ \omega_c$ is $\beta=f/ \tilde \omega
=\omega_c a/2\pi c=(\omegac r_e/2\pi^2 \param c)^{1/3}$, where
$\param=\omegap^2/ \omegac^2$, while $r_e=e^2/4\pi\epsilon_0 m_e
c^2$ is the so-called classical electron radius. In all
experimental situations, one always has $\omegac\ll c/r_e\cong
1.06\times 10^{23}$ Hz (i.e.\ $B\ll 10^{11}$ T), whence $\beta \ll
1$. For instance, for $\param=3$ and $\omegac\approx 10^{12}$ Hz,
which is a typical value for a tokamak, we find $\beta \approx 5
\times 10^{-5}$. It follows that, for all dispersion curves
discussed in section \ref{sw}, for which $\tilde \omega$ was at
most of order unity, one has $f=\beta\tilde\omega\ll 1$. This
implies that, for wavelengths not too much longer than the lattice
parameter, $f$ is negligible with respect to $q$. Hence, with very
good approximation one can operate the substitution
$\widehat{\vect M}({\vect q}, f)\to \widehat{\vect M}({\vect q},
0)$, and it is immediate to see that $\widehat{\vect M}({\vect q},
0)$ is just the matrix $\widehat{\vect M} ({\vect q})$ we
introduced in section \ref{model}. In particular, for $f\to 0$ we
have $M_{ij}^{\rm mac}({\vect k}, \omega) \to -q_i q_j /{\vect
q}^2$ and $\widehat {\vect N}({\vect q}, f) \to \widehat {\vect
N}({\vect q}, 0)=\widehat{\vect N} ({\vect q})$, where
$\widehat{\vect N} ({\vect q})$ is the matrix defined by
(\ref{nij}).  It follows that the dynamical equation
(\ref{clinrad}) reduces to (\ref{clin}) in this approximation. We
have thus verified that the formulas derived by considering purely
coulombian interactions give a fully satisfactory description of
the dispersion relations discussed in section \ref{sw}.

\subsection{The long-wavelength limit}
Let us now examine the form of the dispersion relations in the
long-wavelength limit $\lambda\gg a$. In this case $q\ll 1$, so
that $f$ in general is no longer negligible with respect to $q$.
On the other hand, as we have already observed, in this limit
$\widehat{\vect N}({\vect q}, f)$ can be neglected, so that
(\ref{clinrad}) can be simplified as
\begin{equation}\label{clinlw}
i \frac {\omega e}{m_e}\vect B \times \vect C -\omega^2 \vect C
=\omegap^2 \widehat {\vect M}^{\rm mac}({\vect k}, \omega) \cdot
\vect C \,.
\end{equation}
It is then easy to see that (\ref{clinlw}) leads to the same
dispersion relations as those which are provided by the usual
macroscopic treatment of high frequency waves in a magnetized
plasma.

For instance, let us consider modes with $\vect k$ parallel to
$\vect B$. If $\vect C$ is parallel to $\vect k$, then $\vect B
\times \vect C= 0$ and $\widehat{\vect M}^{\rm mac}({\vect k},
\omega)\cdot \vect C= -\vect C$. It then follows from
(\ref{clinlw}) that longitudinal waves (i.e.\ waves which involve
an oscillation of the electronic density) have frequency $\omega=
\omegap$ independently of the magnetic field $B$ and of the
wavelength $\lambda$, for $\lambda\gg a$. Hence $\omegap$ plays
indeed the role of ``plasma frequency'' also in this model.
Moreover, for transversal waves, i.e.\ $\vect C\cdot \vect k= 0$,
we find that the normal modes allowed by (\ref{clinlw}) correspond
to circularly polarized waves with dispersion relation
\begin{equation}\label{dispgold1}
\frac{c^2k^2}{\omega^2}=1-\frac{\omegap^2}
{\omega(\omega-\omegac)}\,.
\end{equation}
This result exactly coincides with that obtained in the
approximation of MHD at zero temperature, represented by formula
(17.35) and figures 17.4--17.5 of reference \cite{gold}.

Similarly, let us consider the modes with $\vect k$ perpendicular
to $\vect B$ and $k\ll 1/a$. For $\vect C$ parallel to $\vect B$,
we deduce from (\ref{clinlw}) the existence of transversal waves
unaffected by the magnetic field, with dispersion relation
\begin{equation}\label{dispgold2}
c^2k^2=\omega^2-\omegap^2\,.
\end{equation}
These modes correspond to formula (16.32) and figure 16.4 of
\cite{gold}. In addition, for $\vect C\cdot \vect B=0$ we have
modes in which the electrons describe elliptical orbits in planes
orthogonal to $\vect B$. For these modes, (\ref{clinlw}) provides
the dispersion relation
\begin{equation}\label{dispgold3}
\frac{c^2k^2}{\omega^2}=1-\frac{\omegap^2(\omega^2- \omegap^2)}
{\omega^2(\omega^2-\omegap^2 -\omegac^2)}\,,
\end{equation}
which corresponds to formula (17.12) and figure 17.1 of
\cite{gold}.

The behavior of the dispersion curves in the long-wavelength
limit, for $\vect B$ either parallel or perpendicular to $\vect
k$, is shown in the left part of Fig.\ \ref{fig0} for $\omegap^2
/\omegac^2 =3$. In order to compare the scales of the abscissa in
the two graphs of this figure, recall that, for $\omegac\approx
10^{12}$ Hz, we have $\omegac a/2\pi c \approx 5 \times 10^{-5}$,
so that $k c/ \omegac=10$ corresponds to $q=ka/2\pi\approx 5\times
10^{-4}$. This means that the long-wavelength region, represented
in the left graph of Fig.\ \ref{fig0}, appears so narrow in the
right graph that it becomes practically invisible. It is clear
however that the left graph displays the behavior for low $k$ of
the curves of the right graph. It has to be noted in particular
that equation (\ref{clinlw}) is satisfied for $\vect C \cdot \vect
k=0$ and $\omega =0$ independently of $k$, provided $k\ll 1/a$.
Solutions of this type simply correspond to static deformations of
the equilibrium lattice, and represent the limit for $q\ll 1$ of
the lowest solid curve in the right part of Fig.\ \ref{fig0}.

\section{Conclusion}\label{disc}
The main result obtained in this paper is that the classical model
of a neutral plasma (as constituted by point electrons with
Coulomb interactions, moving in a smeared-out positive
background), when linearized about an equilibrium position,
generally presents unstable normal modes. A relevant part of these
modes is stabilized by an external magnetic field only for plasma
densities below a maximal one, which is expressed by a
Brillouin-type formula.

A natural question is then whether  our result, which essentially
refers to a zero temperature situation inasmuch as it deals with
normal modes about an equilibrium configuration, may be
significant also for high temperature plasmas. Indeed, it is well
known that disruptive instabilities occur beyond a density limit
in fusion machines with magnetic confinement \cite{green}. In this
connection one may remark that we are dealing here with an
instability property, and that the raising of temperature tends to
increase disorder rather than creating order. Thus the occurrence
of an instability at zero temperature should imply instability  at
high temperatures as well.

An indication that the instability discussed here for the
linearized system might perhaps be of interest for fusion plasmas,
comes from the remark that the density limit found here turns out
in several cases to be in a fairly good agreement with the limit
empirically encountered in the operation of the tokamaks for
fusion research. For example, for a magnetic field  $B$ = 5 T the
Alcator C-Mod device shows a limit $n_e^M \simeq 3.8 \times
10^{20}$ m$^{-3}$ \cite{labombard}, whereas formula (\ref{dl}),
with $\bparamc\simeq 1.4$ as obtained for a simple cubic lattice,
predicts $n_e^M \simeq 3.4 \times 10^{20}$ m$^{-3}$. Analogously,
at $B$ = 2 T, the DIII-D device \cite{petrie} presents a density
limit of $n_e^M \simeq 6 \times 10^{19}$ m$^{-3}$, which has to be
compared to the prediction $n_e^M \simeq 5.4 \times 10^{19}$
m$^{-3}$ of formula (\ref{dl}).

Although a $B^2$ dependence of the density limit had been noticed
by Granetz \cite{granetz} for the Alcator C experiment, it is
generally believed that the currently available global set of
experimental data on the density limit of toroidal machines is
best fitted by Greenwald's empirical scaling law \cite{green}, according to
which the limit is proportional to the plasma current density in
the tokamak. However, it must be recalled that, despite the large
theoretical work on the subject, at the moment no widely accepted,
first principles model for the density limit in tokamak devices
appears to exist \cite{green}. Thus, the results here presented
might provide a motivation for further experimental
investigations, in order to establish whether a quadratical
dependence on the magnetic field may provide a good description of
the data for at least some class of machines. This might have
relevant implications on the expected performances of future
tokamaks.

\begin{acknowledgments}
This work, supported by the European Communities under the
contract of Association between EURATOM/ENEA, was carried out
within the framework the European Fusion Development Agreement.
\end{acknowledgments}

\appendix

\section{Calculation of the field acting on an electron}\label{appa}
We start from the expression of the charge density within the
plasma
\[
\rho(\vect x, t)= e\sum_{\vect n} \delta^3 (\vect x - \vect
z_{\vect n}(t))+ \rho_{\rm bg}\,.
\]
Up to first order in $\vect c \equiv \vect C \exp(-i\omega t)$, we
can write $\rho= \rho^{(0)} + \rho^{(1)}$, where
\begin{align*}
\rho^{(0)} (\vect x)&= \frac e V -  e\sum_{\vect n} \delta^3
(\vect x - \vect r_{\vect n}) \\
&= -\frac e V \sum_{\vect m \neq \vect 0}\exp (i\vect G_{\vect m}
\cdot \vect x)\,,
\end{align*}
\begin{align*}
\rho^{(1)} (\vect x, t)&= e\sum_{\vect n} \exp(i\vect k \cdot
\vect r_{\vect n})\, \vect c \cdot \mathbf{\nabla}\delta^3
(\vect x - \vect r_{\vect n}) \\
&= i\frac e V \sum_{\vect m}\vect c \cdot (\vect G_{\vect m}+
\vect k)\exp [i(\vect G_{\vect m}+ \vect k) \cdot \vect x]\,.
\end{align*}
In these formulas, the vectors
\begin{equation*}
\vect G_{\vect m}\equiv \frac {2\pi}{V} (m_1 \vect a_2 \times
\vect a_3+ m_2 \vect a_3 \times \vect a_1+ m_3 \vect a_1 \times
\vect a_2)\,,
\end{equation*}
with $\vect m \in \mathbf Z^3$, represent the points of the
reciprocal lattice.

Using the Poisson equation $\Delta \phi =- \rho/\epsilon_0$, we
obtain for the electrostatic potential $\phi$ the corresponding
expansion $\phi= \phi^{(0)}+ \phi^{(1)}$, where
\begin{equation}\label{phi0}
\phi^{(0)} (\vect x)= -\frac e {V\epsilon_0} \sum_{\vect m \neq
\vect 0} \frac {\exp (i\vect G_{\vect m} \cdot \vect x)}{\vect
G_{\vect m}^2}\,,
\end{equation}
\begin{equation*}
\phi^{(1)} (\vect x, t)= \frac {ie} {V\epsilon_0} \sum_{\vect m}
\left.\frac{\vect c \cdot \vect G'_{\vect m}} {|\vect G'_{\vect
m}|^2} \,\exp (i\vect G'_{\vect m} \cdot \vect x)\right|_{\vect
G'_{\vect m}=\vect G_{\vect m}+ \vect k}\,.
\end{equation*}
It follows that the electric field inside the plasma is given in
the dipole approximation by $\vect E= \vect E^{(0)}+ \vect
E^{(1)}$, where
\begin{widetext}
\begin{equation*}
\vect E^{(0)} (\vect x)=- \mathbf{\nabla}\phi^{(0)} (\vect x)=
\frac {ie} {V\epsilon_0} \sum_{\vect m \neq \vect 0} \frac {\vect
G_{\vect m}}{\vect G_{\vect m}^2} \exp (i\vect G_{\vect m} \cdot
\vect x)\,,
\end{equation*}
\begin{equation}\label{etot1}
\vect E^{(1)} (\vect x, t)=- \mathbf{\nabla}\phi^{(1)} (\vect x,t) 
= \frac {e} {V\epsilon_0} \sum_{\vect m} \left.\frac{\vect c \cdot
\vect G'_{\vect m}} {|\vect G'_{\vect m}|^2} \vect G'_{\vect
m}\,\exp (i\vect G'_{\vect m}
\cdot \vect x)\right|_{\vect G'_{\vect m}=\vect G_{\vect m}+ \vect k}\,. 
\end{equation}

The contribution due to the electron $\vect n$ is $\vect E_{\vect
n}= \vect E^{(0)}_{\vect n}+ \vect E^{(1)}_{\vect n}$, where
\begin{equation*}
\vect E^{(0)}_{\vect n} (\vect x)=\frac{e}{4\pi \epsilon_0}
\frac{\vect x -\vect r_{\vect
n}}{|\vect x -\vect r_{\vect n}|^3} 
=\frac {ie} {(2\pi)^3\epsilon_0} \int d^3 \vect p \frac {\vect
p}{p^2} \exp [i\vect p\cdot (\vect x- \vect r_{\vect n})]\,,
\end{equation*}
\begin{equation}\label{e1n}
\vect E^{(1)}_{\vect n} (\vect x, t) 
= \frac{e}{4\pi \epsilon_0} \exp(i \vect k \cdot \vect r_{\vect
n}) \bigg[\frac{\vect c}{|{\vect x}- {\vect r}_{\mathbf{n}}|^3}
-3\frac{{\vect c}\cdot ({\vect x}- {\vect
r}_{\mathbf{n}})}{|{\vect x}- {\vect r}_{\mathbf{n}}|^5} ({\vect
x}- {\vect r}_{\mathbf{n}}) \bigg]
=\frac {e} {(2\pi)^3\epsilon_0} \exp(i \vect k \cdot \vect
r_{\vect n}) \int d^3 \vect p \frac{\vect c \cdot \vect p} {p^2}
\vect p\,\exp [i\vect p \cdot (\vect x- \vect r_{\vect n})]\,.
\end{equation}

Using (\ref{etot1}) and (\ref{e1n}), and proceeding as in
\cite{mcg}, we thus obtain
\begin{align*}
{\vect e}_{\vect n}^{(1)} ({\vect r}_{\vect n}, t)= &\lim_{{\bf
x}\rightarrow {\bf r}_{{\bf n}}} \left[ {\bf E}^{(1)}({\bf
x},t)-{\bf E}_{\bf n}^{(1)} ({\bf x},t)\right]
=\frac {e} {V\epsilon_0} \exp(i \vect k \cdot \vect r_{\vect n})
\lim_{{\bf x}\rightarrow {\bf 0}} \int d^{3}{\bf p}\,e^{i{\bf
p\cdot x}}\frac{\vect p\cdot \vect c}{p^{2}} \vect p
\left[\sum_{{\bf m}}\delta^{3} ({\bf p-H}_{{\bf m}}-{\bf
q})-1\right] \\
=&- \frac {e n_e}{\epsilon_0} \widehat {\vect M}({\vect q}) \cdot
\vect c \exp(i\vect k \cdot \vect r_{\vect n})\,,
\end{align*}
where $\vect q =\vect k a/2\pi$, $\vect H_{\vect m} =\vect
G_{\vect m} a/2\pi$, and
\begin{equation*}
\widehat {\vect M}({\vect q}) = -\lim_{\eta\rightarrow 0^+} \int
d^{3}{\bf p}\, e^{-\eta p^2}c_{ij}(\vect p, \vect q)
\left[ \sum_{{\bf m}}\delta^{3} (\vect p-\vect H_{\bf m})
-1\right]\,.
\end{equation*}
The above integral can be evaluated by expanding the integrand
function $c_{ij}(\vect p, \vect q)$, defined by formula
(\ref{cij}), in powers of $\vect q/p$ about the origin. Denoting
by $c_{ij}^{(k)}$ the term of order $(q/p)^k$, the first four
terms of this expansion are respectively
\begin{align*}
c_{ij}^{(0)}&=\frac{p_{i}p_{j}}{p^{2}}\,, \\
c_{ij}^{(1)}&=\frac{1}{p^{2}}\left( q_{i}p_{j}+q_{j}p_{i}-2{\bf
q}\cdot {\bf p}\frac{p_{i}p_{j}}{p^{2}}\right)\,,\\
c_{ij}^{(2)}&= \frac{1}{p^{2}}\bigg\{ q_{i}q_{j} -2(q_{i}p_{j}
+q_{j}p_{i})\frac{{\bf q}\cdot {\bf p}}{p^{2}} +
\frac{p_{i}p_{j}}{p^{2}}\left[ -{\bf q}^{2} +\frac{4({\bf
q}\cdot {\bf p})^{2}}{p^{2}}\right] \bigg\} \,,\\
c_{ij}^{(3)}&= \frac{1}{p^{4}}\bigg\{ -2{\bf q}\cdot {\bf p}\,
q_{i}q_{j} + (q_{i}p_{j}+q_{j}p_{i})\left[ -{\bf q}^{2}
+\frac{4({\bf q}\cdot {\bf p})^{2}}{p^{2}}\right] 
-4\frac{p_{i}p_{j}}{p^{2}} {\bf q}\cdot {\bf p}\left[ -{\bf q}^{2}
+\frac{2({\bf q}\cdot {\bf p})^{2}}{p^{2}}\right] \bigg\}\,.
\end{align*}
The limit for $\eta \to 0^+$ of the integral of these four terms
gives a polynomial function of $\vect q$ whose coefficients can be
evaluated numerically for any given lattice geometry. The
remainder $\bar c_{ij}\equiv c_{ij}- c_{ij}^{(0)}- c_{ij}^{(1)}-
c_{ij}^{(2)}- c_{ij}^{(3)}$ of the integrand function is of order
$(q/p)^{4}$ for $p \to +\infty$, hence it is possible to put
directly $\eta=0$ before evaluating the integral. This procedure
leads for an isotropic lattice to formulas (\ref{e1}) and
(\ref{mij})--(\ref{nij}), with
\begin{align}
\alpha &=\lim_{\eta \rightarrow 0^{+}}\left(-\sum_{{\bf m}\neq
{\bf 0}}\frac{\exp \left( -\eta {\bf H}_{{\bf m}}^{2}
\right)}{{\bf H}_{{\bf m}}^{2}} +\frac{2\pi ^{3/2}}{\sqrt{\eta }}
\right)\,, \label{alpha}\\
\beta &=\lim_{\eta \rightarrow 0^{+}}\left(-\sum_{{\bf m}\neq {\bf
0}}\frac{\left({\bf H}_{{\bf m}}\right)_1^4}{\left|{\bf H}_{{\bf
m}}\right|^{6}} \exp \left( -\eta {\bf H}_{{\bf m}}^{2} \right)
+\frac{2\pi ^{3/2}}{5\sqrt{\eta }} \right)\,.\label{beta}
\end{align}

Note that for an isotropic lattice one can also write
\[
\sum_{\vect m \neq 0} \bar c_{ij}(\vect H_{\vect m}, \vect q)=
\lim_{L \to +\infty}\sum_{\vect m \neq 0, |\vect m|\leq L} \tilde
c_{ij}(\vect H_{\vect m}, \vect q)\,,
\]
where
\begin{equation*}
\tilde c_{ij}(\vect H, \vect q)= \frac{(H_{i} +q_{i}) (H_{j}
+q_{j})} {({\bf H}+ {\bf q})^{2}} -\frac {\delta_{ij}}3 -\frac
{q_{i}q_{j}+{\bf q}^{2}\delta_{ij}/3-2q_{i}^{2} \delta_{ij}}
{{\vect H}^2} +\frac {2H_1^4}{H^6} (2q_{i}q_{j}+{\bf q}^{2}
\delta_{ij}-5q_{i}^{2} \delta_{ij})\,.
\end{equation*}
\end{widetext}

\section{Calculation of the electrostatic energy at equilibrium}\label{appb}

From formula (\ref{u1}) it follows that the electrostatic energy
of our model of plasma in its equilibrium configuration is
\[
U_0= -\frac{Ne}2 \lim_{\vect x\to \vect 0}\left[\phi^{(0)}(\vect
x)+ \frac e{4\pi \epsilon_0 x} \right]\,,
\]
where $\phi^{(0)}$ is the potential generated by all the charges
of the plasma, given by (\ref{phi0}), while $-e/4\pi\epsilon_0 x$
is the potential generated by the electron at the origin.

In order to numerically compute $U_0$, it is convenient to
introduce the auxiliary electrostatic potential $\psi$ generated
by an array of charge distributions, each given by the
superposition of a point charge $-e$ and a gaussian of total
charge $+e$. This can formally be written as
\[
\psi(\vect x)= \sum_{\vect n}\bar \psi (\vect x- \vect r_{\vect
n}) \,,
\]
where $\vect r_{\vect n}$ are the points of the Bravais lattice,
and the function $\bar \psi$ satisfies
\begin{equation}\label{barpsi}
\Delta\bar\psi (\vect x)= \frac e{\epsilon_0}\left[\delta^3(\vect
x) -\left(\frac \eta \pi\right)^{3/2} \exp(-\eta x^2)\right]\,,
\end{equation}
$\eta$ being an arbitrary parameter. We have $U_0=A+B$, where
\[
A= -\frac{Ne}2 \lim_{\vect x\to \vect 0}\left[\psi(\vect x)+ \frac
e{4\pi \epsilon_0 x} \right]
\]
and
\[
B= -\frac{Ne}2 \lim_{\vect x\to \vect 0}\left[\phi^{(0)}(\vect x)
-\psi(\vect x) \right]\,.
\]
We are now going to show that both terms $A$ and $B$ can be
evaluated as the sums of rapidly convergent series.

A standard integration of (\ref{barpsi}) provides
\[
\bar\psi (\vect x)= -\frac {eF(\sqrt \eta x)}{4\pi \epsilon_0
x}\,,
\]
where the function $F$ is defined by (\ref{funf}). We have
\begin{align*}
\lim_{\vect x\to \vect 0}\left[\bar\psi(\vect x)+ \frac e{4\pi
\epsilon_0 x} \right]&= \frac e{4\pi \epsilon_0} \lim_{x\to 0}
\frac{1-F(\sqrt \eta x)}x \\
&= \frac{e\sqrt\eta}{2\epsilon_0 \pi^{3/2}} \,,
\end{align*}
whence
\begin{equation}\label{a}
A=-\frac{Ne^2}{8\pi \epsilon_0}\left[2\sqrt{\frac \eta\pi}
-\sum_{\vect n \neq \vect 0}\frac{F(\sqrt{\eta} |\vect r_{\vect
n}|)}{|\vect r_{\vect n}|}\right]\,.
\end{equation}


In order to calculate $B$, we observe that, due to its lattice
periodicity, $\psi$ can be Fourier expanded as
\begin{equation}\label{psi}
\psi(\vect x)= \sum_{\vect m} \tilde\psi_{\vect m}\exp(i\vect
G_{\vect m}\cdot \vect x)\,,
\end{equation}
where $\vect G_{\vect m}$ are the points of the reciprocal
lattice, and
\begin{equation*}
\tilde\psi_{\vect m}=\frac 1V\int \bar\psi(\vect x)\exp(-i\vect
G_{\vect m} \cdot \vect x) d^3 \vect x \,.
\end{equation*}
We have
\begin{align*}
\tilde\psi_{\vect 0}=&-\frac e{\epsilon_0 V}\int_0^{+\infty}
F(\sqrt
\eta x) x\,dx \\
=&-\frac e{\epsilon_0 V} \sqrt{\frac \eta\pi}\int_0^{+\infty}
e^{-\eta x^2} x^2\,dx= -\frac e{4\epsilon_0 V\eta}
\end{align*}
and, for $\vect m\neq \vect 0$,
\begin{align*}
\tilde\psi_{\vect m}=&-\frac e{\epsilon_0 V |\vect G_{\vect
m}|}\int_0^{+\infty} F(\sqrt \eta x) \sin(|\vect G_{\vect
m}|x)dx \\
=&-\frac e{\epsilon_0 V |\vect G_{\vect m}|^2}\left[1-\sqrt{\frac
\eta\pi}\int_{-\infty}^{+\infty} e^{-\eta x^2} \cos(|\vect
G_{\vect m}|x)dx \right]\\
=&-\frac e{\epsilon_0 V |\vect G_{\vect
m}|^2}\left[1-\exp\left(-\frac{|\vect G_{\vect m}|^2}{4\eta}
\right)\right]\,.
\end{align*}
From (\ref{phi0}) and (\ref{psi}) it then follows that
\begin{align}\label{b}
B&=\frac{Ne}2\left[\tilde\psi_{\vect 0} +\sum_{\vect m \neq \vect
0} \left(\tilde\psi_{\vect m}-\frac e{\epsilon_0 V |\vect G_{\vect
m}|^2}\right)\right] \nonumber \\
&=-\frac{Ne^2}{2\epsilon_0 V}\left[\frac 1{4\eta}- \sum_{\vect m
\neq \vect 0}\frac 1{|\vect G_{\vect m}|^2}\exp\left(-\frac{|\vect
G_{\vect m}|^2}{4\eta} \right)\right] \,. 
\end{align}
By putting $\xi=a^2 \eta/\pi$, from (\ref{a}) and (\ref{b}) one
finally obtains (\ref{u0}), with $b$ given by (\ref{mu}).

\section{Calculation of $\paramc(\theta)$}\label{theta}
The equation ${\rm det}\, \widehat {\vect A} (\vect k, \omega)
=0$, with $\widehat {\vect A} (\vect k, \omega)$ given by
(\ref{amat}) and $\vect q= (1/2) \vect u_3$ , can be explicitly
written as
\begin{align}\label{pu}
&u^3- (1+\param)u^2 -\param[\param Z(3Z+2)+ Z-
\xi]u \nonumber\\
&-\param^3 Z^2(1+2Z) =0\,,
\end{align}
with $u=\omegar^2=\omega^2/\omegac^2$, $Z\equiv \bar N(1/2)\cong
0.05212$, and $\xi \equiv (1+3Z) \cos^2 \theta$. We see that the
left-hand side of this equation is a third-degree polynomial in
$u$, which we shall call $P(u)$. Hence the corresponding normal
modes will all be stable (i.e.\ have a real frequency) provided
this polynomial admits three real nonnegative roots. We first note
that $P(0)<0$ for all $\param>0$. A necessary condition for the
existence of three positive roots is then $P'(0)>0$, where $P'$
denotes the derivative of $P$. Hence we must have
\begin{equation}\label{ppu}
\param<\frac{\xi-Z}{Z(3Z+2)}\,,
\end{equation}
which implies in particular $\xi>Z$, i.e.\ $\cos\theta >
\sqrt{Z/(1+3Z)} \cong 0.2123$. Whenever (\ref{ppu}) is satisfied,
it is easily seen that $P'(u)$ has two positive roots, which we
shall call $u_1$ and $u_2$, with $0<u_1<u_2$. Then $P(u)$ will
have three positive roots if and only if $P(u_1)>0$ and
$P(u_2)<0$. With some simple algebra, one sees that the validity
of both these conditions is equivalent to the single inequality
\begin{align}\label{diseq}
&4\xi Z(1+3Z)^3 \param^3 -[Z^2(1+3Z)^2 -2Z \xi (18 Z^2+ 21Z +5)
\nonumber\\
&+(36Z^2+ 24Z+1) \xi^2]\param^2 -2[Z^2(1+Z) -Z(4+9Z)\xi \nonumber\\
&+(1+6Z)\xi^2 -2\xi^3] \param
- Z^2+2Z\xi -\xi^2<0 \,.
\end{align}
It is found that, for $\xi>Z$, the third degree polynomial in
$\param$ on the left-hand side of (\ref{diseq}) has a single
positive root $\bar \param$, and that this root satisfies
(\ref{ppu}). Putting $\paramc(\theta)=\bar\param$, it then follows
that (\ref{pu}) admits three real nonnegative roots for all
$\param$ such that $0\leq\param\leq \paramc(\theta)$.


\begin{thebibliography}{}


\bibitem{brill} L. Brillouin, \emph{Phys.
Rev.} \textbf{67}, 260 (1945).

\bibitem{davidson} R. C. Davidson, \emph{Physics of Nonneutral
Plasmas} (Addison-Wesley, Redwood City, 1990).

\bibitem{bg1} D. Bohm and E. P. Gross, \emph{Phys. Rev.}
\textbf{75}, 1851 and 1864 (1949).

\bibitem{bp1} D. Bohm and D. Pines, \emph{Phys. Rev.}
\textbf{82}, 625 (1951).

\bibitem{pb2} D. Pines and D. Bohm, \emph{Phys. Rev.}
\textbf{85}, 338 (1952).

\bibitem{langmuir} I. Langmuir,
\emph{Proc. Nat. Acad. Sci.} \textbf{14}, 627 (1928).

\bibitem{tonks} L. Tonks and I. Langmuir, \emph{Phys. Rev.}
\textbf{33}, 195 (1929).

\bibitem{Jackson} J. D.~Jackson, \emph{Classical Electrodynamics}
(John Wiley \& Sons, New York, 1975).

\bibitem{mcg} M. Marino, A. Carati and L. Galgani, \emph{Ann. Phys.}
\textbf{322}, 799 (2007).

\bibitem{kittel} C. Kittel, \emph{Introduction to Solid State
Physics}, eighth edition (John Wiley \& Sons, Hoboken, 2005).

\bibitem{dirac} P.A.M. Dirac, \emph{Proc. R. Soc. London}
\textbf{167}, 148 (1938).

\bibitem{marino} M. Marino, \emph{Ann. Phys.} \textbf{301}, 85 (2002).

\bibitem{cg} A. Carati and L. Galgani,
\emph{Nuovo Cimento} \textbf{118 B}, 839 (2003).

\bibitem{gold} R. J. Goldston and P.H. Rutherford, \emph{Introduction to
Plasma Physics} (IOP Publishing, Bristol, 1995).

\bibitem{green} M. Greenwald,
\emph{Plasma Phys. Control. Fusion} \textbf{44}, R27 (2002).

\bibitem{labombard} B. LaBombard et al., \emph{Phys. Plasmas} \textbf{8}, 2107 (2001).

\bibitem{petrie} T. W. Petrie, A. G. Kellmann and M. Ali Mahdavi,
\emph{Nucl. Fusion} \textbf{33}, 929 (1993).

\bibitem{granetz} R. S. Granetz, \emph{Phys. Rev. Lett.}
\textbf{49}, 658 (1982).

\end{thebibliography}
\end{document}